\documentstyle[amssymb,aps]{revtex}

\begin{document}
\title{Convex rigid cover method in studies of quantum pure-states of many
continuous variables}
\author{Zai-Zhe Zhong}
\address{Department of Physics, Liaoning Normal University, Dalian 116029, Liaoning,\\
China. E-mail: zhongzaizheh@hotmail.com}
\maketitle

\begin{abstract}
In this paper we prove that every pure-state $\Psi ^{\left( N\right) }$ of N
(N$\geqslant 3)$ continuous variables corresponds to a pair of convex rigid
covers (CRCs) structures in the continuous-dimensional Hilbert-Schmidt
space. Next we strictly define what are the partial separability and
ordinary separability, and discuss how to use CRCs to describe various
separability. We discuss the problem of the classification of $\Psi ^{\left(
N\right) }$ and give a  kinematical explanation of the local unitary
operations acting upon $\Psi ^{\left( N\right) }$. Thirdly, we discuss the
invariants of classes and give a possible physical explanation.

PACC numbers: 03.67.Mn, 03.65.Ud, 03.67.Hk
\end{abstract}

In quantum mechanics and quantum information, the{\em \ }study of the
multipartite quantum systems{\em \ }is more difficult. For instance, for the
general multipartite quantum systems the problems of the criteria of various
separability, of the entanglement measures, and of the classification and
invariants, etc. all are not solved better as yet. As for the cases of many
continuous variables (MCVs), the problems are even more complex, because,
generally, the various related expressions of the quantum states of MCVs are
more complex than once of the quantum states of discrete variables.
Especially, the study of the entanglement problems(e.g. see [1,2]) of the
quantum states of MCVs, for instance, the discussion of the quantum
teleportation, even if in the simplest case of two CVs[3], is still more
complex than ordinary qubit system. In order to study the multipartite
systems, in [4] we have we suggested an effective method, i.e. the `method
of convex rigid frames (CRFs)' which is a non-traditional way. However the
method of CRFs only can be used in the cases of the quantum states of
discrete variables (quNit pure-states). In this paper we consider the more
important cases, i.e. we give a way in the study of entanglement problem of
the quantum states of MCVs, we call it the `method of convex rigid covers
(CRCs)' (see below) which is also a non-traditional way, of course. As a
result of the complexness of the problems, it is not only a simple
continuous-dimensional generalization of the method of CRFs, especially the
cases of various separability, related calculations, etc.

In this paper, first we explain what are the CRCs and prove that every
pure-state $\Psi ^{\left( N\right) }$ of N continuous variables (CVs) always
corresponds to a pair of CRCs in the continuous-dimensional Hilbert-Schmidt $%
\left( \text{H-S}\right) $ space[5] (about it, see below). Next we strictly
what are the partial separability (relate the cases of discrete variables,
see [6,7]) and ordinary separability of pure-states of MCVs, and how we use
the method of CRCs to describe these separability. Further we discuss the
problem of the classification of $\Psi ^{\left( N\right) }$ and give an
explanation of kinematics of the local unitary transformation (LU)[8,9]
acting upon $\Psi ^{\left( N\right) }$. Thirdly, we discuss the invariants
of classes and give a possible physical explanation, and a concrete example
is given..

In the first place, we introduce what is a continuous-dimensional H-S space,
in this paper we only consider the following cases. Let $H_i\equiv H\left(
\mu _i\right) \left( i=1,\cdots ,m\right) $ be the ordinary Hilbert space
consisting of complex value functions of variable $\mu _i.$ For the sake of
simpleness$,$ we assume that the definition field of m CVs $\left( \mu
\right) _m\equiv \left( \mu _1,\cdots ,\mu _m\right) $ is the real field $%
{\Bbb R}^m.$ If the basis of $H_i$ has been chosen as $\mid B\left( \mu
_i\right) $ 
\mbox{$>$}%
, where B$\left( \mu _i\right) \equiv $B$\left( \mu _i,{\bf x}\right) $ (in
the following the space coordinates ${\bf x,}$ generally, are omitted) are
basic vector obeying 
\begin{equation}
<B\left( \mu _i^{\prime }\right) \mid B\left( \mu _i\right) >=\delta \left(
\mu _i^{\prime }-\mu _i\right) \text{, }\int d\mu _id\mu _i^{\prime }\mid
B\left( \mu _i\right) ><B\left( \mu _i^{\prime }\right) \mid =1
\end{equation}
Now we consider the space $H=H_1\otimes \cdots H_m,$ the basis can be chose
as $\left\{ \mid B_{\left( \mu \right) _m}>\right\} \equiv \left\{ \mid
B\left( \mu _1\right) >\otimes \cdots \otimes \mid B\left( \mu _m\right)
>\right\} $ 
\begin{equation}
<B_{\left( \mu ^{\prime }\right) _m}\mid B_{\left( \mu \right) _m}>=\delta
\left( \left( \mu ^{\prime }\right) -\left( \mu \right) \right) ,\int \left(
d\mu \right) ^m\mid B_{\left( \mu \right) _m}><B_{\left( \mu \right) _m}\mid
=1
\end{equation}
where the $\delta $ function $\delta \left( \left( \mu ^{\prime }\right)
-\left( \mu \right) \right) \equiv \delta \left( \mu _1^{\prime }-\mu
_1\right) \cdots \delta \left( \mu _m^{\prime }-\mu _m\right) $ and $\left(
d\mu \right) ^m\equiv d\mu _1\cdots d\mu _m$. Then a normal pure-state $\mid
\Psi ^{\left( N\right) }>$, whose continuous `coordinates' are $\left\{
c_{\left( \mu \right) _m}\right\} ,$ can be expressed as 
\begin{equation}
\mid \Psi ^{\left( m\right) }>=\int \left( d\mu \right) ^mc_{\left( \mu
\right) _m}\mid B_{\left( \mu \right) _m}>,c_{\left( \mu \right) _m}\equiv
<B_{\left( \mu \right) _m}\mid \Psi ^{\left( m\right) }>,\int \left( d\mu
\right) ^m\left| c_{\left( \mu \right) _m}\right| ^2=1
\end{equation}
Now a Hermitian operator $L$ can be represent by a continuous-dimensional
matrix whose matrix entries are 
\begin{equation}
\left[ L\right] _{\left( \mu ^{\prime }\right) _m,\left( \mu \right)
_m}=\int \left( d{\bf x}\right) ^3<B_{\left( \mu ^{\prime }\right) _m}\left( 
{\bf x}\right) \mid L\mid B_{\left( \mu \right) _m}\left( {\bf x}\right) >
\end{equation}
The set of all operators $L$ form a linear space ${\cal L}^{\left(
m+m\right) }$ on the real number field, a $L$ can be called a vector or a
point in ${\cal L}^{\left( m+m\right) }$. In ${\cal L}^{\left( m+m\right) }$
if we define the interior product of vectors $A$ and $B$ is defined as 
\[
<A,B>=\int \left( d\mu \right) ^m\left( A^{\dagger }B\right) _{\left( \mu
\right) _m,\left( \mu \right) _m}=\int \int \left( d\mu \right) ^m\left(
d\nu \right) ^m\left[ A\right] _{\left( \mu \right) _m,\left( \nu \right)
_m}\left[ B\right] _{\left( \mu \right) _m,\left( \nu \right) _m}^{*} 
\]
where the star denotes the complex conjugation. The modulus (the length) of
a vector $A$ is defined by $\left\| A\right\| =\sqrt{<A,A>}$ =$\sqrt{\int
\int \left( d\mu \right) ^m\left( d\nu \right) ^m\left| \left[ A\right]
_{\left( \mu \right) _m,\left( \tau \right) _m}\right| ^2}$, and the
distance between two points $A\;$and $B$ is defined as $d\left( A,B\right)
=\left\| A-B\right\| .$ It can easily be verified that ${\cal L}^{\left(
m+m\right) }$ forms a distance space, which is the continuous-dimensional
H-S space. The density operator (matrix) $\rho $ is a vector (a point) of $%
{\cal L}^{\left( m+m\right) }$, $\rho ^{\left( m\right) }=\mid \Psi ^{\left(
m\right) }><\Psi ^{\left( m\right) }\mid ,\;\left[ \rho ^{\left( m\right)
}\right] _{\left( \mu ^{\prime }\right) _m,\left( \mu \right) _m}=c_{\left(
\mu ^{\prime }\right) _m}c_{\left( \mu \right) _m}^{*}$. In this paper, all
operators discussed are density operators.

Notice that every pure-state $\rho ^{\left( m\right) }=\mid \Psi ^{\left(
m\right) }><\Psi ^{\left( m\right) }\mid $\ is a vertex in ${\cal L}^{\left(
m+m\right) }.$\ Now, we assume that $\rho ^{\left( m\right) }=\rho ^{\left(
m\right) }\left( \xi \right) _n\equiv \rho ^{\left( m\right) }\left( \xi
_1,\cdots ,\xi _n\right) $\ yet is dependent on other n real variables $%
\left( \xi \right) _n\in {\Bbb R}^n,$then every $\rho ^{\left( m\right)
}\left( \xi \right) _n$\ is a vertex in ${\cal L}^{\left( m+m\right) }.$\ If
we take $\rho ^{\left( m\right) }\left( \xi _1,\cdots ,\xi _n\right) $ as a
mapping that $\rho ^{\left( m\right) }\left( \xi \right) _n:\;{\Bbb R}%
^n\longrightarrow {\cal L}^{\left( m+m\right) },$ then the image of $\rho
^{\left( m\right) }\left( \xi \right) _n$, $\sigma ^{\left( n\right) }=\rho
^{\left( m\right) }\left( {\Bbb R}^n\right) $, \ is a piece of a curve
surface of a convex body (the maximal dimension of the cover $\sigma
^{\left( n\right) }$ is n)$,$ we call $\sigma ^{\left( n\right) }$\ a
`convex cover' which just can be described by the density matrix $\rho
^{\left( m\right) }\left( \xi \right) _n$. Next, suppose that $\lambda
\left( \xi \right) _n\equiv \lambda \left( \xi _1,\cdots ,\xi _n\right) $\
is a nonnegative real function obeying $\lambda \left( \xi \right)
_n\geqslant 0,\;\int \left( d\xi \right) ^n\lambda \left( \xi \right) _n=1,$
then we obtain a mixed-state$\;p^{\left( n\right) }=\int \left( d\xi \right)
^n\lambda \left( \xi \right) _n\rho ^{\left( m\right) }\left( \xi \right) _n$
(we assume that it is convergent)$,$\ which is a point $p^{\left( n\right) }$%
\ in ${\cal L}^{\left( m+m\right) }$\ and near the convex cover $\sigma
^{\left( n\right) }$\ (the point $p^{\left( n\right) }$\ is `covered 'by $%
\sigma ^{\left( n\right) }).$\ We can take a figure of speech: If $\lambda
\left( \xi \right) _n$\ is the density of mass distributing on $\sigma
^{\left( n\right) }$, the total mass is 1, $\rho ^{\left( m\right) }\left(
\xi \right) _n$ corresponds the `coordinates' of a point in $\sigma ^{\left(
n\right) }$, then $p^{\left( n\right) }$\ just is the `centre of mass' of $%
\sigma ^{\left( n\right) }$. Obviously $p^{\left( n\right) }$\ is only
determined by $\lambda \left( \xi \right) _n.$ In this paper${\em ,}$ every
related $\sigma ^{\left( n\right) }$\ and the corresponding point p$^{\left(
m\right) }{\em \ }$discussed only can be moved in ${\cal L}^{\left(
m+m\right) }$ as a rigid body as in the classical mechanics, so we call them
a `convex rigid cover(CRC)', simply read it the symbol $CRC_{\left( \xi
\right) _n}^{\left( m\right) }=\left\{ \rho ^{\left( m\right) }\left( \xi
\right) _n,\lambda ^{\left( m\right) }\left( \xi \right) _n\right\} $.

{\it Definition 1}. Two convex rigid covers $CRC_{\left( \xi \right)
_n}^{\left( m\right) }=\left\{ \rho ^{\left( m\right) }\left( \xi \right)
_n,\lambda ^{\left( m\right) }\left( \xi \right) _n\right\} $\ and $%
CRC_{\left( \xi \right) _n}^{\prime \left( m\right) }=\left\{ \rho ^{\prime
\left( m\right) }\left( \xi \right) _n,\lambda \left( \xi \right) _n\right\} 
$\ are called to be identical, if and only if $d\left( \rho ^{\left(
m\right) }\left( \xi \right) _n,\rho ^{\left( m\right) }\left( \xi ^{\prime
}\right) _n\right) =d\left( \rho ^{\prime \left( m\right) }\left( \xi
\right) _n,\rho ^{\prime \left( m\right) }\left( \xi ^{\prime }\right)
_n\right) $\ and $\lambda ^{\left( m\right) }\left( \xi \right) _n=\lambda
^{\left( m\right) \prime }\left( \xi ^{\prime }\right) _n$\ for any $\left(
\xi \right) _n,\left( \xi ^{\prime }\right) _n$.\ In this case we call the
process $CRC_{\left( \xi \right) _n}^{\left( m\right) }\longrightarrow
CRC_{\left( \xi \right) _n}^{\prime \left( m\right) }$\ a `motion from $%
CRC_{\left( \xi \right) _n}^{\left( m\right) }$\ to $CRC_{\left( \xi \right)
_n}^{\prime \left( m\right) }$'.

Obviously, this\ identical relation is an equivalence relation, so all CRCs
can be classified by using of this relation.

Now we consider N (N$\geqslant 3)$\ particles of MCVs, suppose that $\left\{
\mid B_{\left( \mu \right) _N}>\right\} $ has been chosen. In the following, 
${\Bbb Z}_M$\ denotes the integer set ${\Bbb Z}_M=\left\{ 1,\cdots
,M\right\} .$\ If two ordered proper subsets $\left( r\right) _P\equiv
\left\{ r_1,\cdots ,r_P\right\} \left( \;1\leqslant r_1<\cdots <r_P\leqslant
M\right) $\ and $\left( s\right) _{M-P}\equiv \left\{ s_1,\cdots
,s_{M-P}\right\} \left( 1\leqslant s_1<\cdots <s_{M-P}\leqslant M\right) $\
in ${\Bbb Z}_M$\ obey $\left( r\right) _P\cup \left( s\right) _{M-P}={\Bbb Z}%
_M,\;\left( r\right) _P\cap \left( s\right) _{M-P}=\emptyset ,$ where $P$\
is an integer, 1$\leqslant P\leqslant M-1,$\ then the set $\left\{ \left(
r\right) _P,\left( s\right) _{M-P}\right\} $\ forms a partition of ${\Bbb Z}%
_M$, we denote it by the symbol $\left( r\right) _P\Vert \left( s\right)
_{M-P}$. For a given partition $\left( r\right) _P\Vert \left( s\right) _Q$,
we define an orthogonal complete basis $\left\{ \mid B_{\left( r\right)
_P\Vert \left( s\right) _Q}>\right\} $\ (i.e. only the order in $\left( \mu
\right) _N$ is changed) by{\em \ } 
\begin{eqnarray}
\left( B_{\left( r\right) _P\Vert \left( s\right) _Q}\right) &=&\mid
B_{\left( \mu \right) _{\left( r\right) _P}}>\otimes \mid B_{\left( \mu
\right) _{\left( s\right) _Q}}> \\
&\mid &B_{\left( \mu \right) _{\left( r\right) _P}}>\equiv \mid B\left( \mu
_{r_1}\right) >\otimes \cdots \otimes \mid B\left( \mu _{r_P}\right) >,\mid
B_{\left( \mu \right) _{\left( s\right) _Q}}>\equiv \mid B\left( \mu
_{s_1}\right) >\otimes \cdots \otimes \mid B\left( \mu _{s_Q}\right) > 
\nonumber
\end{eqnarray}
We denote the state $\mid \Psi _{\left( r\right) _P\Vert \left( s\right)
_Q}^{\left( N\right) }$%
\mbox{$>$}%
corresponding to $\mid \Psi ^{\left( N\right) }$%
\mbox{$>$}%
by 
\begin{eqnarray}
&\mid &\Psi _{\left( r\right) _P\Vert \left( s\right) _Q}^{\left( N\right)
}>=\int \left( d\mu \right) _{\left( r\right) _P}^P\int \left( d\mu \right)
_{\left( s\right) _Q}^Q\widetilde{c}_{\left( \left( \mu \right) _{\left(
r\right) _P},\left( \mu \right) _{\left( s\right) _Q}\right) }\mid \left(
B_{\left( r\right) _P\Vert \left( s\right) _Q}\right) >  \nonumber \\
&=&\int \left( d\mu \right) _{\left( s\right) _Q}^Q\left[ \int \left( d\mu
\right) _{\left( r\right) _P}^P\widetilde{c}_{\left( \left( \mu \right)
_{\left( r\right) _P},\left( \mu \right) _{\left( s\right) _Q}\right) }\mid
B_{\left( \mu \right) _{\left( r\right) _P}}>\right] \mid B_{\left( \mu
\right) _{\left( s\right) _Q}}>
\end{eqnarray}
where we simply write $\widetilde{c}_{\left( \left( \mu \right) _{\left(
r\right) _P},\left( \mu \right) _{\left( s\right) _Q}\right) }\equiv $ $%
\widetilde{c}_{\left( \mu _{r_1},\cdots ,\mu _{r_P},\mu _{s_1},\cdots ,\mu
_{s_Q}\right) }=c_{\left( \mu \right) _N},$ and $\left( \mu \right)
_N=\left( \mu \right) _{\left( r\right) _P}\cup \left( \mu \right) _{\left(
s\right) _Q}$ but it is in the natural order. Obviously, $\mid \Psi _{\left(
r\right) _P\Vert \left( s\right) _Q}^{\left( N\right) }>$ and\ $\mid \Psi
^{\left( N\right) }>$ are completely the same in the physics. Further, we
write 
\begin{eqnarray}
&\mid &\Psi _{\left( r\right) _P\Vert \left( s\right) _Q}^{\left( N\right)
}>=\int \left( d\mu \right) _{\left( s\right) _Q}^Q\eta \left( \left( \mu
\right) _{\left( s\right) _Q}\right) \mid \varphi \left( \left( \mu \right)
_{\left( s\right) _Q}\right) >\mid B_{\left( \mu \right) _{\left( s\right)
_Q}}>  \nonumber \\
\eta \left( \left( \mu \right) _{\left( s\right) _Q}\right) &=&\sqrt{\int
\left( d\mu \right) _{\left( r\right) _P}^P\left| c_{\left( \mu \right)
_N}\right| ^2},\mid \varphi \left( \left( \mu \right) _{\left( s\right)
_Q}\right) >=\int \left( d\mu \right) _{\left( r\right) _P}^Pe_{\left(
r\right) _P}^P\left( \left( \mu \right) _{\left( s\right) _Q}\right) \mid
B_{\left( \mu \right) _{\left( r\right) _P}}>  \nonumber \\
e_{\left( r\right) _P}^P\left( \mu \right) _{\left( s\right) _Q} &=&\frac{%
c_{\left( \mu \right) _N}}{\sqrt{\int \left( d\mu \right) _{\left( r\right)
_P}^P\left| c_{\left( \mu \right) _N}\right| ^2}}
\end{eqnarray}
where $\left( \mu \right) _{\left( s\right) _Q}$ play the roles of Q
`parameters' (as $\left( \xi \right) _n$ as in the above), and it is easily
verified that $\mid \varphi \left( \left( \mu \right) _{\left( s\right)
_Q}\right) >$ is a normal, and $\int \left( d\mu \right) _{\left( s\right)
_Q}^Q\eta ^2\left( \left( \mu \right) _{\left( s\right) _Q}\right) =1$.
Therefore corresponding to a $\mid \Psi ^{\left( N\right) }>$ and a $\left(
r\right) _P\Vert \left( s\right) _Q$, we at once point out a CRC as 
\begin{eqnarray}
CRC_{\left( \mu \right) _{\left( s\right) _Q}}\left( \rho ^{\left( N\right)
}\right) &=&\left\{ \rho ^{\left( P\right) }\left( \rho ^{\left( N\right)
},\left( \mu \right) _{\left( s\right) _Q}\right) ,\lambda \left( \rho
^{\left( N\right) },\left( \mu \right) _{\left( s\right) _Q}\right) \right\} 
\text{ (}\left( \mu \right) _{\left( s\right) _Q}\text{ can changes}) \\
\rho ^{\left( P\right) }\left( \rho ^{\left( N\right) },\left( \mu \right)
_{\left( s\right) _Q}\right) &=&\mid \varphi \left( \left( \mu \right)
_{\left( s\right) _Q}\right) ><\varphi \left( \left( \mu \right) _{\left(
s\right) _Q}\right) \mid ,\;\lambda \left( \rho ^{\left( N\right) },\left(
\mu \right) _{\left( s\right) _Q}\right) =\eta ^2\left( \left( \mu \right)
_{\left( s\right) _Q}\right) =\int \left( d\mu \right) _{\left( r\right)
_P}^P\left| c_{\left( \mu \right) _N}\right| ^2  \nonumber
\end{eqnarray}
Symmetrically, if we consider $\mid \Psi _{\left( s\right) _Q\Vert \left(
r\right) _P}^{\left( N\right) }>$, then we obtain 
\begin{eqnarray}
CRC_{\left( \mu \right) _{\left( r\right) _P}}\left( \rho ^{\left( N\right)
}\right) &=&\left\{ \rho ^{\left( Q\right) }\left( \rho ^{\left( N\right)
},\left( \mu \right) _{\left( r\right) _P}\right) ,\lambda \left( \rho
^{\left( N\right) },\left( \mu \right) _{\left( r\right) _P}\right) \right\} 
\text{ (}\left( \mu \right) _{\left( r\right) _P}\text{ can changes}) 
\nonumber \\
\rho ^{\left( Q\right) }\left( \rho ^{\left( N\right) },\left( \mu \right)
_{\left( r\right) _P}\right) &=&\mid \varphi \left( \left( \mu \right)
_{\left( r\right) _P}\right) ><\varphi \left( \left( \mu \right) _{\left(
r\right) _P}\right) ,\mid \varphi \left( \left( \mu \right) _{\left(
r\right) _P}\right) >=\int \left( d\mu \right) _{\left( s\right) _Q}^Q\frac{%
c_{\left( \mu \right) _N}}{\sqrt{\int \left( d\mu \right) _{\left( s\right)
_Q}^Q\left| c_{\left( \mu \right) _N}\right| ^2}}\mid B_{\left( \mu \right)
_{\left( s\right) _Q}}> \\
\lambda \left( \rho ^{\left( N\right) },\left( \mu \right) _{\left( r\right)
_P}\right) &=&\eta ^2\left( \left( \mu \right) _{\left( r\right) _P}\right)
=\int \left( d\mu \right) _{\left( s\right) _Q}^Q\left| c_{\left( \mu
\right) _N}\right| ^2  \nonumber
\end{eqnarray}
Therefore for a $\mid \Psi ^{\left( N\right) }>$ and a given $\left(
r\right) _P\Vert \left( s\right) _Q,$ we always write a pair of CRCs as 
\begin{equation}
CRC_{\left( r\right) _P\Vert \left( s\right) _Q}\left( \rho ^{\left(
N\right) }\right) =\left[ CRC_{\left( \mu \right) _{\left( s\right)
_Q}}\left( \rho ^{\left( N\right) }\right) ,\;CRC_{\left( \mu \right)
_{\left( r\right) _P}}\left( \rho ^{\left( N\right) }\right) \right]
\end{equation}
\ 

Now, we discuss the separability problems by using the above way. First we
need to strictly define the various separability of a pure-state $\mid \Psi
^{\left( N\right) }>$ of MCVs$.$

{\it Definition 2.}{\em \ }The pure-state $\mid \Psi ^{\left( N\right) }>$\
is called to be $\left( r\right) _P-\left( s\right) _Q$-separable, if $\mid
\Psi ^{\left( N\right) }>$\ can be decomposed as{\em \ } 
\begin{eqnarray}
&\mid &\Psi _{\left( r\right) _P\Vert \left( s\right) _Q}^{\left( N\right)
}>=\mid \Psi _{\left( r\right) _P}^{\left( P\right) }>\mid \Psi _{\left(
s\right) _Q}^{\left( Q\right) }>,\text{or }\rho _{\left( r\right) _P\Vert
\left( s\right) _{M-P}}^{\left( N\right) }=\rho _{\left( r\right)
_P}^{\left( P\right) }\otimes \rho _{\left( s\right) _{M-P}}^{\left(
Q\right) }\equiv \mid \Psi _{\left( r\right) _P}^{\left( P\right) }><\Psi
_{\left( r\right) _P}^{\left( P\right) }\mid \otimes \mid \Psi _{\left(
s\right) _Q}^{\left( Q\right) }><\Psi _{\left( s\right) _Q}^{\left( Q\right)
}\mid  \nonumber \\
&\mid &\Psi _{\left( r\right) _P}^{\left( P\right) }>=\int \left( d\mu
\right) _{\left( r\right) _P}^Pe_{\left( \mu \right) _{\left( r\right)
_P}{}}\mid B_{\left( \mu \right) _{\left( r\right) _P}}^P>,\mid \Psi
_{\left( s\right) _Q}^{\left( Q\right) }>=\int \left( d\mu \right) _{\left(
s\right) _Q}^Qf_{\left( \mu \right) _{\left( s\right) _Q}}\mid B_{\left( \mu
\right) _{\left( s\right) _Q}}^Q>
\end{eqnarray}
this means that there is the relation as $c_{\mu _1\cdots \mu _N}=e_{\left(
\mu \right) _{\left( r\right) _P}{}}\times f_{\left( \mu \right) _{\left(
s\right) _Q}}$. The above separability is a partial separability. If $\mid
\Psi ^{\left( N\right) }>$\ \ is not $\left( r\right) _P-\left( s\right)
_{M-P}$-separable, then we call it is $\left( r\right) _P-\left( s\right)
_{M-P}$-inseparable.

{\it Definition 3.} The pure-state $\mid \Psi ^{\left( N\right) }>$\ is
called to be separable (disentangled), if {\em \ } 
\begin{equation}
\mid \Psi ^{\left( N\right) }>=\prod_{i=1}^N\left( \int d\mu _ie_{\mu
_i}\mid B\left( \mu _i\right) >\right)
\end{equation}
Therefore the separability, in fact, is a `full separability'. It is easily
verified that $\mid \Psi ^{\left( N\right) }>$\ is separable, if and only if
it always partially separable with respect to all possible partition. The
following theorem is a main result in this paper, it , in fact, is a
criterion of the partial separability.

{\it Theorem 1}{\em .\ }The pure-state $\rho ^{\left( N\right) }=\mid \Psi
^{\left( N\right) }><\Psi ^{\left( N\right) }\mid $\ is $\left( r\right)
_P-\left( s\right) _{M-P}$-separable if and only if the cover $CRC_{\left(
\mu \right) _{\left( s\right) _Q}}\left( \rho ^{\left( N\right) }\right) $
(or $CRC_{\left( \mu \right) _{\left( r\right) _P}}\left( \rho ^{\left(
N\right) }\right) $) shrinks to one point (pure-state vertex), i.e. all $%
d\left( \rho ^{\left( P\right) }\left( \rho ^{\left( N\right) },\left( \mu
\right) _{\left( s\right) _Q}\right) -\rho ^{\left( P\right) }\left( \rho
^{\left( N\right) },\left( \mu ^{\prime }\right) _{\left( s\right)
_Q}\right) \right) =0$\ $($or all $d\left( \rho ^{\left( Q\right) }\left(
\rho ^{\left( N\right) },\left( \mu \right) _{\left( r\right) _P}\right)
-\rho ^{\left( Q\right) }\left( \rho ^{\left( N\right) },\left( \mu \right)
_{\left( r\right) _P}\right) \right) =0)$\ for any $\left( \mu \right)
_{\left( s\right) _Q},\left( \mu ^{\prime }\right) _{\left( s\right) _Q}$ $%
\left( \text{or }\left( \mu \right) _{\left( r\right) _P},\left( \mu
^{\prime }\right) _{\left( r\right) _P}\right) $.

{\bf Proof}. Suppose that the pure-state $\rho ^{\left( N\right) }=\mid \Psi
^{\left( N\right) }><\Psi ^{\left( N\right) }\mid $\ is $\left( r\right)
_P-\left( s\right) _{M-P}$-separable, i.e. Eq.(11) holds. Then from Eq..(7), 
$\rho ^{\left( P\right) }\left( \rho ^{\left( N\right) },\left( \mu \right)
_{\left( s\right) _Q}\right) $, in fact, obviously is independent on $\left(
\mu \right) _{\left( r\right) _P},$ this means that it will shrink to a
fixed point $\rho ^{\left( P\right) },$ i.e. cover $CRC_{\left( \mu \right)
_{\left( s\right) _Q}}\left( \rho ^{\left( N\right) }\right) $\ indeed
shrink a point. Similarly, for $CRC_{\left( \mu \right) _{\left( r\right)
_P}}\left( \rho ^{\left( N\right) }\right) .$

Conversely, if $CRC_{\left( r\right) _P}\left( \rho ^{\left( N\right)
}\right) $\ shrink to a point $\sigma =\mid \varphi ><\varphi \mid ,$\ this
require that $e_{\left( \mu \right) _{\left( s\right) _Q}}\left( \left( \mu
\right) _{\left( r\right) _P}\right) $ in Eq.(7), in fact, is independent on 
$\left( \mu \right) _{\left( r\right) _P}$,\ then Eq.(7) leads to that there
are $e_{\left( \mu \right) _{\left( r\right) _P}}$\ and $f_{\left( \mu
\right) _{\left( s\right) _Q}}$\ such that $c_{\left( \mu \right)
_N}=e_{\left( \mu \right) _{\left( r\right) _P}}f_{\left( \mu \right)
_{\left( s\right) _Q}}$ and $\mid \Psi _{\left( r\right) _P\Vert \left(
s\right) _{M-P}}>=\mid \varphi >\otimes \mid \psi >,$\ where $\mid \psi
>=\int \left( d\mu \right) _{\left( s\right) _Q}^Qf_{\left( \mu \right)
_{\left( s\right) _Q}}\mid B_{\left( \mu \right) _{\left( s\right) _Q}}>,$ $%
\rho ^{\left( N\right) }$\ is $\left( r\right) _P-\left( s\right) _{M-P}$%
-separable. $\square $

{\it Corollary}. For a pure-state $\rho ^{\left( N\right) }$\ of MCVs and
the partition $\left( r\right) _P\Vert \left( s\right) _Q,$\ $CRC_{\left(
\mu \right) _{\left( s\right) _Q}}\left( \rho ^{\left( N\right) }\right) $\
and $CRC_{\left( \mu \right) _{\left( r\right) _P}}\left( \rho ^{\left(
N\right) }\right) $\ both shrink to points, or both not.

The proofs are evident from the proof of the Theorem 1.

Therefore in view of method of CRCs in this paper, every separable
(disentangled) pure-state of MCVs is an extremely special state, i.e. of
which all CRCs with respect to any partition must be shrunk into a point.
Conversely, if $\rho ^{\left( N\right) }$\ is inseparable with respect to
any one of a partition, then it must be entangled.

Secondly, by the method of CRCs we discuss the `kinematical explanation' of
the LU acting upon the pure-states of MCVs.

{\it Definition 4}{\em . }We call two pure-states $\rho ^{\left( N\right) }$%
\ and $\rho ^{\prime \left( N\right) }$\ of N CVs are `equivalent by
motion', symbolize by $\rho ^{\left( N\right) }\backsim \rho ^{\prime \left(
N\right) }$, if and only if $CRC_{\left( \mu \right) _{\left( s\right)
_Q}}\left( \rho ^{\left( N\right) }\right) $\ and $CRC_{\left( \mu \right)
_{\left( r\right) _P}}\left( \rho ^{\left( N\right) }\right) $\ are
identical (see the Definition 1) with respect to all possible non-null
proper subset $\left( r\right) _P\left( 1\leqslant P\leqslant M-1\right) $.

{\it Corollary}{\em .\ }If two pure-states $\rho ^{\left( N\right) }$\ and $%
\rho ^{\prime \left( N\right) }$\ of N CVs\ are equivalent by motion, then $%
\rho ^{\left( N\right) }$\ and $\rho ^{\prime \left( N\right) }$\ both are $%
\left( r\right) _P-\left( s\right) _{M-P}$-separable (or both $\left(
r\right) _P-\left( s\right) _{M-P}$-inseparable), with respect to any $%
\left( r\right) _P\Vert \left( s\right) _{M-P},\;$i.e. the partial
separability and separability are invariants of the equivalence classes by
motion.

From the definitions 1 and 4, the proof is obvious.

The following theorem is just a kinematical explanation of LU.

{\it Theorem 2}. Two pure-states $\rho ^{\left( N\right) }$ and $\rho
^{\prime \left( N\right) }$\ are equivalent by motion (see the Definition
4), if and only if there are N unitary transformations $u\left( \mu
_i\right) \left( i=1,\cdots ,N\right) $ that ${\it t}$%
\begin{equation}
\rho ^{\prime \left( N\right) }=u\left( \mu _1\right) \otimes \cdots \otimes
u\left( \mu _N\right) \rho ^{(N)}u^{\dagger }\left( \mu _N\right) \otimes
\cdots \otimes u^{\dagger }\left( \mu _1\right)
\end{equation}

{\bf Proof. }Notice that in the H-S space the unitary transformations and
only the unitary transformations just can keep the invariance of distances
and modulus of the vectors, by this fact, the proof is obvious. $\square $

Thirdly, we discuss the invariant problem of classes. Obviously all
invariants of $CRC_{\left( \mu \right) _{\left( s\right) _Q}}\left( \rho
^{\left( N\right) }\right) $ (and $CRC_{\left( \mu \right) _{\left( r\right)
_P}}\left( \rho ^{\left( N\right) }\right) )$ of motions just are the
invariants of LU of a pure-state $\rho ^{\left( N\right) },$ for instance,
the `areas' of convex cover $\sigma ^{\left( P\right) }$ in $CRC_{\left( \mu
\right) _{\left( r\right) _P}}\left( \rho ^{\left( N\right) }\right) $($%
\sigma ^{\left( P\right) }$ is the image of the mapping $\rho ^{\left(
Q\right) }\left( \rho ^{\left( N\right) },\left( \mu \right) _{\left(
r\right) _P}\right) $ from ${\Bbb R}^P$ to ${\cal L}^{\left( Q+Q\right) },$
and similarly for $\sigma ^{\left( Q\right) }$ )$,$ the curvatures in every
point in $\sigma ^{\left( P\right) }$ (and in $\sigma ^{\left( Q\right)
}),\cdots ,$ (of course, the `total mass' (always is 1), the `density of
mass distribution' $\lambda \left( \rho ^{\left( N\right) },\left( \mu
\right) _{\left( r\right) _P}\right) \left( \text{or }\lambda \left( \rho
^{\left( N\right) },\left( \mu \right) _{\left( s\right) _Q}\right) \right) $
is invariant). In addition, there are those invariants describing convexity,
etc. At present, we cannot yet understand what are the meaning in quantum
information for the most of them. However, the `areas' have a quite natural
explanation as follows.

In order to avoid the infinity and to simplify calculation, we can choose a
finite region $\Omega ^N\subset {\Bbb R}^N$ , and assume that $\left( \mu
\right) _{\left( r\right) _P}$ and $\left( \mu \right) _{\left( s\right) _Q}$%
, respectively, full in $\Omega _{\left( r\right) _P}$ and $\Omega _{\left(
s\right) _Q}$, $\Omega _{\left( r\right) _P}\times \Omega _{\left( s\right)
_Q}=\Omega ^N$ . Let $S_{\Omega _{\left( r\right) _P}}\left( \sigma ^{\left(
P\right) }\right) $ denote the `area' (here, the word `area', in fact,
represents a measure of some dimension, and we assume that it is finite) of
the convex cover $\sigma ^{\left( P\right) }$ on $\Omega _{\left( r\right)
_P}$. Similarly, $S_{\Omega _{\left( s\right) _Q}}\left( \sigma ^{\left(
Q\right) }\right) .$ Now we denote the pair 
\begin{equation}
S_{\Omega ^N}\left( \rho ^{\left( N\right) }\right) =\left\{ S_{\Omega
_{\left( r\right) _P}}\left( \sigma ^{\left( P\right) }\right) ,S_{\Omega
_{\left( s\right) _Q}}\left( \sigma ^{\left( Q\right) }\right) \right\}
\end{equation}
Obviously, $S_{\Omega ^N}\left( \rho ^{\left( N\right) }\right) $ is an
invariant under motions (LUs) of $\rho ^{\left( N\right) }$. From the
Theorem 1, its corollary and the fact that a convex cover shrinks to a point
if and only if its area vanishes for any $\Omega ^N$, then we know that $%
\rho $ is $\left( r\right) _P-\left( s\right) _Q$-separable if and only if $%
S_{\Omega ^N}\left( \rho ^{\left( N\right) }\right) =\left( 0,0\right) $ for
any $\Omega ^N.$ Conversely, if $S_{\Omega ^N}\left( \rho ^{\left( N\right)
}\right) \neq \left( 0,0\right) $ for some $\Omega ^N$, then $\rho ^{\left(
N\right) }$ is $\left( r\right) _P-\left( s\right) _Q$-inseparable, where
the value of $S_{\Omega _{\left( r\right) _P}}\left( \sigma ^{\left(
P\right) }\right) $means that `on $\Omega _{\left( r\right) _P}$ the degree
of the difficulty of the factor $\rho _{\left( r\right) _P}^{\left( N\right)
}$ to be separated out from $\rho ^{\left( N\right) }$'$.$ Similarly, for $%
S_{\Omega _{\left( s\right) _Q}}\left( \sigma ^{\left( P\right) }\right) .$
Therefore we can regard that $S_{\Omega ^N}\left( \rho ^{\left( N\right)
}\right) $ denotes some `on $\Omega ^N$ the degree of the measure of the $%
\left( r\right) _P-\left( s\right) _Q$-inseparability'. It is quite
interesting that, generally, $S_{\Omega _{\left( r\right) _P}}\left( \sigma
^{\left( P\right) }\right) \neq S_{\Omega _{\left( s\right) _Q}}\left(
\sigma ^{\left( Q\right) }\right) $ unless they both vanish.

However, generally the calculations of $S_{\Omega _{\left( r\right)
_P}}\left( \sigma ^{\left( P\right) }\right) $\ or $S_{\Omega _{\left(
s\right) _Q}}\left( \sigma ^{\left( Q\right) }\right) $\ both are difficult,
even if in the most ideal case that $\sigma ^{\left( P\right) }$\ and $%
\sigma ^{\left( Q\right) }$\ are differentiable manifolds (an useful way,
see [10] and its references). Here, we only discuss a special case that the
partion is $\left( r\right) _P\Vert \left( s\right) _Q=\left( 1,\cdots
,P\right) \Vert \left( P+1,\cdots ,N\right) ,$ and the `supper-rectangular
parallelepipeds' $\Omega _{\left( r\right) _P}\equiv $\ $\left(
a_{r_1},b_{r_1}\right) \times \cdots \times \left( a_{r_P},b_{r_P}\right) $, 
$\Omega _{\left( s\right) _Q}\equiv $\ $\left( a_{s_1},b_{s_1}\right) \times
\cdots \times \left( a_{s_Q},b_{s_Q}\right) $ and $\Omega ^N=\left(
a_1,b_1\right) \times \cdots \times \left( a_N,b_N\right) ,$ where $a_i$ and 
$b_i$ are real, $a_i<b_i$ for $i=1,\cdots ,N.$\ Then $\sigma ^{\left(
P\right) },$\ generally, is P-dimensional in the above H-S space 
\begin{eqnarray}
\sigma ^{\left( P\right) } &=&\left\{ \rho ^{\left( Q\right) }\left( \mu
_1,\cdots ,\mu _P\right) \mid :\rho ^{\left( Q\right) }\left( \mu _1,\cdots
,\mu _P\right) =\mid \Phi ^{\left( Q\right) }\left( \mu _1,\cdots ,\mu
_P\right) ><\Phi ^{\left( Q\right) }\left( \mu _1,\cdots ,\mu _P\right) \mid
\right\} \\
&\mid &\Phi ^{\left( Q\right) }\left( \mu _1,\cdots ,\mu _P\right) >=\int
d\mu _{P+1}\cdots d\mu _Ne_{\left( \mu _{P+1},\cdots ,\mu _N\right) }\left(
\mu _1,\cdots ,\mu _P\right) \mid B_{\left( \mu _{P+1},\cdots ,\mu _N\right)
}>  \nonumber \\
e_{\left( \mu _{P+1},\cdots ,\mu _N\right) }\left( \mu _1,\cdots ,\mu
_P\right) &=&\frac{c_{\left( \mu \right) _N}}{\sqrt{\int d\mu _{P+1}\cdots
d\mu _N\left| c_{\left( \mu \right) _N}\right| ^2}}
\end{eqnarray}
Thus we can calculate the `area' of curve cover $\sigma ^{\left( P\right) }$
on $\Omega _{\left( r\right) _P}$ as 
\begin{eqnarray}
S_{\Omega _{\left( r\right) _P}}\left( \sigma ^{\left( P\right) }\right)
&=&\int_{\Omega _{\left( r\right) _P}}\left\| d\rho ^{\left( N-1\right)
}\left( \mu _1\right) \right\| \\
&=&\int_{a_1}^{b_1}d\mu _1\cdots \int_{a_P}^{b_P}d\mu _{r_P}\sqrt{\int
\left( d\mu \right) ^Q\left( d\mu ^{\prime }\right) ^Q\left| \frac \partial {%
\partial \mu _1}\cdots \frac \partial {\partial \mu _P}\left( \frac{\left|
c_{\left( \mu \right) _N}c_{\left( \overline{\mu ^{\prime }}\right)
_N}^{*}\right| ^2}{\int \left( d\mu \right) ^Q\left| c_{\left( \mu \right)
_N}\right| ^2\int \left( d\mu ^{\prime }\right) ^Q\left| c_{\left( \overline{%
\mu ^{\prime }}\right) _N}\right| ^2}\right) \right| }  \nonumber
\end{eqnarray}
where $\left( d\mu \right) ^Q\equiv d\mu _{P+1}\cdots d\mu _N,\left( d\mu
^{\prime }\right) ^Q\equiv d\mu _{P+1}^{\prime }\cdots d\mu _N^{\prime
},\left( \overline{\mu ^{\prime }}\right) _N\equiv \left( \overline{\mu
^{\prime }}_1,\cdots ,\overline{\mu ^{\prime }}_N\right) ,$ $\overline{\mu
^{\prime }}_i{}=\mu _{i\text{ }}$ for 1$\leqslant i\leqslant P,$ and $%
\overline{\mu ^{\prime }}_i{}=\mu _{i\text{ }}^{\prime }$ for $P+1\leqslant
i\leqslant N$. Similarly, we can calculate $S_{\left( s\right) _Q}\left(
\sigma ^{\left( Q\right) }\right) $ in this case. It is obvious that $\mid
\Psi ^{\left( N\right) }>$ is $\left( r\right) _P-\left( s\right) _Q$%
-separable if and only if $S_{\Omega _{\left( r\right) _P}}\left( \sigma
^{\left( P\right) }\right) \equiv 0$ for any $\left( a_i,b_i\right) $.

What a pity, as it has been pointed out as in [4] that $S_{\Omega ^N}\left(
\rho ^{\left( N\right) }\right) $ $\left( \text{and }S_{\Omega _{\left(
r\right) _P}}\left( \sigma ^{\left( P\right) }\right) \right) $cannot be
taken as a measure of the partial entanglement.

{\bf Discussions. }If in the system there are both discrete variables and
MCVs , then we can similarly give a method in which there are some mixture
of CRFs and CRCs, and some corresponding conclusions still hold.

{\bf Conclusion. }In the study of{\bf \ }pure-state $\Psi ^{\left( N\right)
} $ of MCVs, the method of CRCs is effective. First by using of this method
we can discuss the separability (partial separability and ordinary
separability). Next, by this method the set of all $\Psi ^{\left( N\right) }$
can be classified, and we find that the LU acting upon a $\Psi ^{\left(
N\right) }$ can be explanted as a motion of CRCs, thus all motion invariants
are LU invariants. Especially, an possible physical explanation of the pair
of the areas of CRCs. is that they represent the degree of the measure of
the $\left( r\right) _P-\left( s\right) _Q$-inseparability on some region $%
\Omega ^N.$

\end{document}